\documentclass[a4paper]{jpconf}
\usepackage{amsmath}
\usepackage{amssymb}
\usepackage{bm}
\usepackage{graphicx}
\begin{document}
\title{QMC study of the chiral Heisenberg Gross-Neveu universality class}

\author{%
Yuichi Otsuka$^{1,2}$,
Kazuhiro Seki$^{2}$,
Sandro Sorella$^{1,3,4}$,
and Seiji Yunoki$^{1,2,5,6}$
}

\address{%
$^1$Computational Materials Science Research Team, 
RIKEN Center for Computational Science (R-CCS), 
Kobe, Hyogo 650-0047, Japan\\
$^2$Quantum Computational Science Research Team, 
RIKEN Center for Quantum Computing (RQC), 
Saitama 351-0198, Japan\\
$^3$SISSA -- International School for Advanced Studies, 
Via Bonomea 265, 34136 Trieste, Italy\\
$^4$Democritos Simulation Center CNR--IOM Instituto Officina 
dei Materiali, Via Bonomea 265, 34136 Trieste, Italy\\
$^5$Computational Quantum Matter Research Team, 
RIKEN Center for Emergent Matter Science (CEMS), 
Wako, Saitama 351-0198, Japan\\
$^6$Computational Condensed Matter Physics Laboratory,
RIKEN,
Wako, Saitama 351-0198, Japan
}

\ead{otsukay@riken.jp}

\begin{abstract}
We investigate a quantum criticality of an antiferromagnetic phase transition 
in the Hubbard model on a square lattice with a $d$-wave pairing field
by large-scale auxiliary-field quantum Monte Carlo simulations.
Since the $d$-wave pairing filed induces Dirac cones in the non-interacting 
single-particle spectrum, the quantum criticality should correspond to the
chiral Heisenberg universality class in terms of the Gross-Neveu theory, which
is the same as those expected in the Hubbard model on the honeycomb lattice,
despite the unit cells being different (e.g.,  they contain one and two sites, 
respectively). We show that both the two phase transitions, expected to occur 
on the square and on the honeycomb lattices, indeed have the same quantum 
criticality. 
We also argue that details of the models, i.e., the way of counting the total
number $N$ of fermion components and the anisotropy of the Dirac cones, 
do not change the critical exponents. 
\end{abstract}

\section{Introduction}

Accurate calculation of critical exponents of phase transitions 
to identify universality classes is one of the major issues in 
computational physics, especially for Monte Carlo simulations.
There are many successful examples found in the studies of 
classical~\cite{Peczak1991,Holm1993,Campostrini2002} and 
quantum~\cite{Wang2006,Wenzel2009,Shao2016,Ma2018}
spin systems.
Recently, another family of the quantum phase transitions that involve 
fermionic degrees of freedom has attracted much attention in this context
because sign-problem-free quantum Monte Carlo (QMC) methods as well as modern 
analytical techniques such as functional renormalization-group approach
are applicable to investigate the fermionic quantum criticality.

The most well-studied example would be the Hubbard model on the honeycomb lattice
(hereinafter referred to as the honeycomb lattice model),
of which the antiferromagnetic (AF) semimetal(SM)-insulator phase transition was 
examined from the view point of 
the Mott transition~\cite{Sorella1992,Paiva2005,Meng2010,Sorella2012,Otsuka2013}.
Later, since the connection between the effective theory of the honeycomb lattice model
and the Gross-Neveu (GN) model in high-energy physics was pointed out~\cite{Herbut2006,Assaad2013}, 
the focus has been shifted from the model-dependent quantum phase transitions
to the universal nature of the quantum criticality 
because the existence of the universality classes for the interacting Dirac fermions
had been well formulated by the GN model~\cite{Rosenstein1993}.
For the AF transition, which corresponds to the chiral-Heisenberg universality class
in terms of the GN theory, reliable estimations of the critical exponents have been 
obtained~\cite{ParisenToldin2015,Otsuka2016}
employing the honeycomb lattice model
and the Hubbard model on the square lattice with $\pi$-flux~\cite{Otsuka2002,Otsuka2014} 
(referred to as the $\pi$-flux model in the following).

In this paper, we revisit the chiral-Heisenberg universality class 
using another lattice model, 
the square-lattice Hubbard model with a $d$-wave pairing field~\cite{Otsuka2020a},
which we call the d-SC model.
If we consider brick-wall square lattices~\cite{Ixert2014,Hatsugai2006}, it can be seen
that the honeycomb lattice model and the $\pi$-flux model are smoothly connected 
to each other even without assuming the universality class.
On the other hand, the present d-SC model is quite different from these two
models as discussed in the next section, while the low-energy effective model
is the same for all the three models, i.e., the GN model breaking the $SU(2)$ symmetry.
Therefore, the examination of the critical exponents for the d-SC model
is expected to serve as an independent and nontrivial check of the previous estimations.

\section{Model and method}

The Hamiltonian of the d-SC model reads as follows:

\begin{equation}
 H = H_{\mathrm{BCS}} + H_{U}, \label{eq:model}
\end{equation}
where
\begin{equation}
 H_{\mathrm{BCS}}  
 = 
\sum_{\langle i, j \rangle}
 \left\{
 \begin{pmatrix}
	c_{i \uparrow}^{\dagger} & 	c_{i \downarrow}^{}	
 \end{pmatrix}
 \begin{pmatrix}
  -t_{} & \Delta_{ij} \\
 \Delta_{ij}^{\ast} & t_{}
 \end{pmatrix}
 \begin{pmatrix}
  c_{j \uparrow}^{} \\
  c_{j \downarrow}^{\dagger}	
 \end{pmatrix} 
 + \mathrm{h.c}
 \right\} \label{eq:H_BCS} \\
\end{equation}
and
\begin{equation}
H_{U}  = U \sum_{i} n_{i \uparrow} n_{i \downarrow}. \label{eq:Hubbard}
\end{equation}
Here, $t$ is the transfer integral between the nearest sites, 
chosen as an energy unit ($t$=1), and 
$\Delta_{ij}$ denotes the $d$-wave pairing field with its amplitude being
uniform, i.e., $|\Delta_{ij}|=\Delta$.
We consider the model at half filling.
The Hubbard interaction denoted by $U$ triggers the AF transition
at the strong coupling regime, where the mass gap opens.
In the noninteracting limit ($U=0$) for finite $\Delta$, 
the ground-state is the SM phase having four Dirac points at the Fermi 
level in the momentum space.
This state is different from those of the honeycomb lattice or 
the $\pi$-flux model which has the two Dirac cones.
In addition, the unit cell of the d-SC model has one site,
whereas the honeycomb lattice or the $\pi$-flux model has two sublattices.
Thus, at the level of the lattice model, they are indeed different.
However, in the low energy continuum limit, these models should be described
by the GN model with the same number of the fermion components, $N=8$, 
with the spin degrees of freedom considered.
This is why we expect the same universality class for the different lattice models.
Furthermore, the d-SC model has another unique feature that 
an anisotropy of the Dirac cone can be tuned by changing $\Delta$.
It is isotropic as in the case of the honeycomb lattice or the $\pi$-flux model 
only at $\Delta=1$, and otherwise the velocity at the Dirac point depends on 
the direction in the momentum space. 
Taking advantage of this feature, we also study whether the anisotropy affects 
the quantum criticality.

Since the square lattice is bipartite and the particle-hole symmetry holds at
half filling, we investigate the d-SC model by the auxiliary-field QMC 
method without facing the negative-sign problem~\cite{Blankenbecler_PRD1981,Hirsch_PRB1985,White_PRB1989}.
The ground-state expectation value of a physical observable $O$ is evaluated
by projection from a left (right) trial wave function $\langle \psi_{\textrm{L}}|$
($|\psi_{\textrm{R}}\rangle$) as 
\begin{equation}
 \langle O \rangle =
  \frac{
  \langle \psi_{\mathrm{L}}| e^{-\frac{\tau}{2} H} 
  O
  e^{-\frac{\tau}{2} H}  |\psi_{\mathrm{R}} \rangle
  }{
  \langle \psi_{\mathrm{L}}| e^{-\tau H} |\psi_{\mathrm{R}} \rangle
  },
\end{equation}
where $\tau$ is projection time and is divided by the Suzuki-Trotter decomposition into
$\tau/M=\Delta \tau$ with $M$ being integer.
We set $\tau$ to be proportional to linear dimension of the square lattice $L$
and chose $\Delta \tau=0.1$ to reduce the systematic errors compared to the stochastic errors.
The simulations are performed on finite-size clusters of
$L$=8, 12, 16, 20, 24, 32, 40 with periodic boundary conditions
for the isotropic ($\Delta=1.0$) and anisotropic ($\Delta=0.5$) cases.
Therefore, the Dirac points at $(\frac{\pi}{2},\frac{\pi}{2})$ and other symmetry equivalent momenta are allowed in all these 
clusters.

\section{Results}

We calculate the spin structure factor 
$S(\bm{k})=L^{-2}\sum_{i,j}e^{i\bm{k}(\bm{r}_{i}-\bm{r}_{j})}\langle \bm{S}_{i}\cdot\bm{S}_{j}\rangle$
in the standard notation and 
the quasiparticle weight 
$Z(U,L)=D_{\sigma}(U,L)/D_{\sigma}(0,L)$
estimated from the equal-time Green's function
$D_{\sigma}(U,L)=L^{-2}\sum_{i}\langle c_{j \sigma}^{\dagger}c_{i \sigma} \rangle$
at the maximum distance~\cite{Seki2019}.
Then, the obtained data are analyzed; 
by a conventional method to first extrapolate the results to the thermodynamics limit ($1/L\rightarrow 0$);
by a more sophisticated method called the crossing-point analysis~\cite{Shao2016}; and
by a rather involved method of data collapse.

\begin{figure}
\centering
 \includegraphics[width=0.24\textwidth]{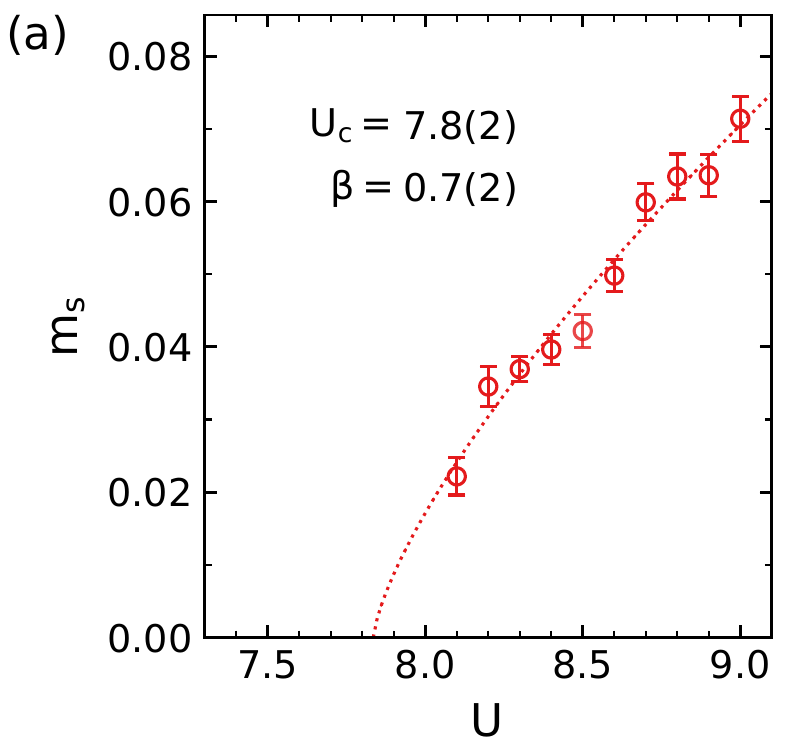}
 \includegraphics[width=0.24\textwidth]{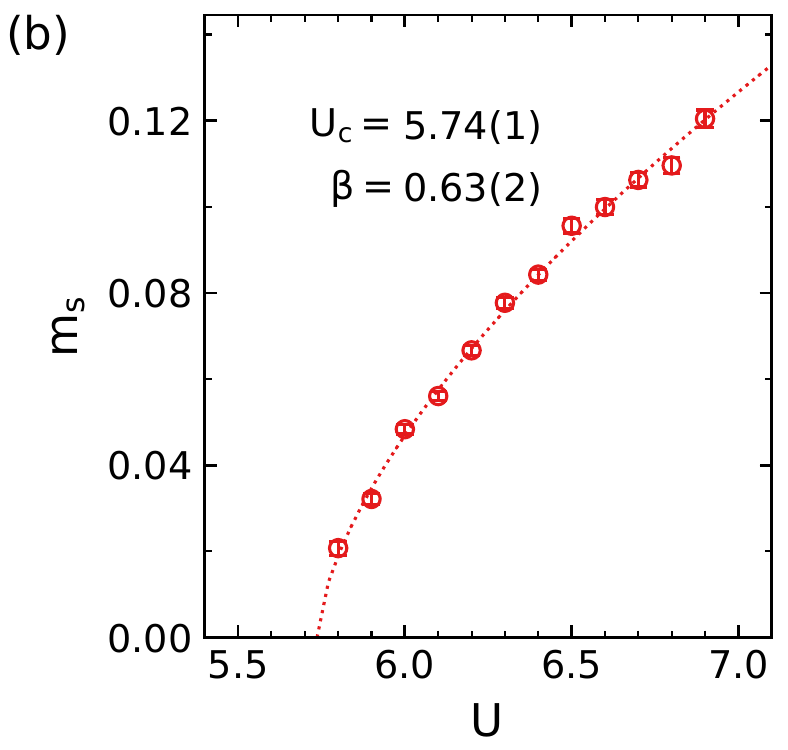}
 \includegraphics[width=0.24\textwidth]{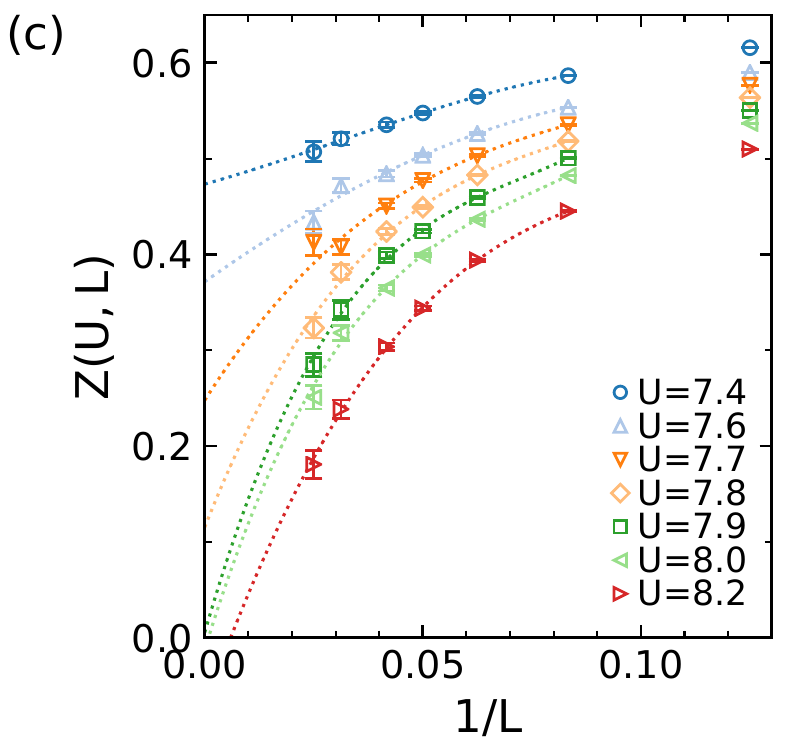}
 \includegraphics[width=0.24\textwidth]{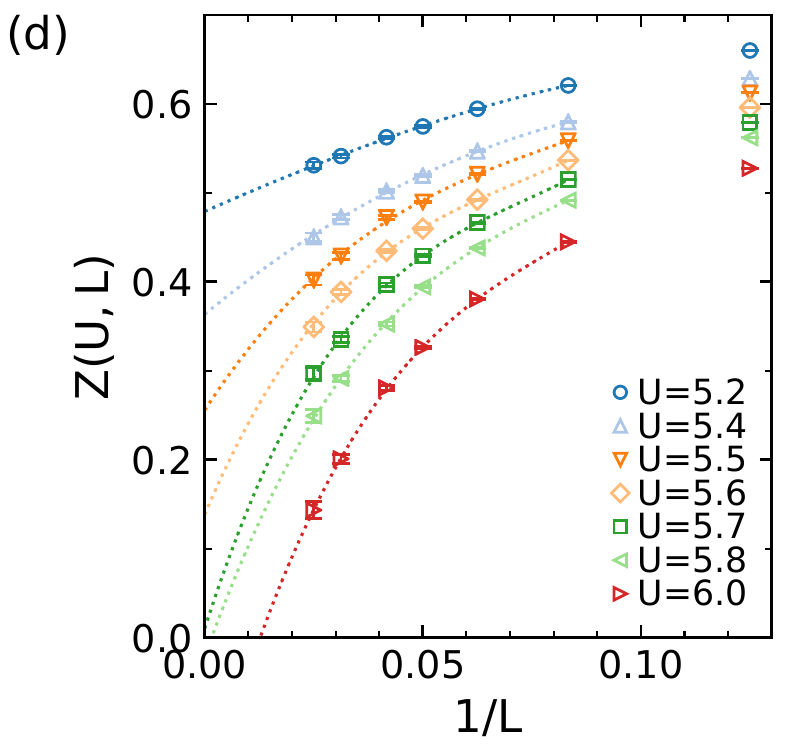}
\caption{\label{fig:poorman}%
$U$-dependence of the AF order parameter 
for (a) $\Delta=1.0$ and (b) $\Delta=0.5$,
and $1/L$-extrapolation of the quasiparticle weight
for (c) $\Delta=1.0$ and (d) $\Delta=0.5$.
}
\end{figure}
The critical point $U_{\textrm{c}}$ dividing the SM and the AF insulator
is in principle obtained as the value of $U$ at which the order 
parameter sets in.
In Figs.~\ref{fig:poorman}(a) and \ref{fig:poorman}(b), we plot the 
staggered magnetization, that is, the AF order parameter, calculated as
$m_{s}=\lim_{1/L \to 0}\sqrt{S(\pi, \pi)}/L$
as a function of $U$.
From these plots, the critical points  and the critical 
exponent $\beta$ are estimated by fitting with the critical behavior 
$m_{s} \sim (U/U_{\textrm{c}}-1)^{\beta}$ as
$U_{\textrm{c}}$=7.8(2) [5.74(1)] and $\beta$=0.7(2) [0.63(2)]
for $\Delta$=1.0 (0.5).
Since the statistical errors are larger for the strong-coupling region,
the error bars for $\Delta$=1.0 are large, 
which makes it difficult to safely conclude that the exponents are consistent
between $\Delta$=1.0 and 0.5 or between the d-SC model and the honeycomb
lattice or the $\pi$-flux model.
We also find large error bars for the quasiparticle weight as shown in
Figs.~\ref{fig:poorman}(c) and \ref{fig:poorman}(d).
In addition, it is not obvious which fitting function is suited 
both for the SM and the insulating phase
to extrapolate $Z(U,L)$ to the thermodynamics limit $1/L=0$ at each $U$.
We are thus not able to plot the quasiparticle weight as the function of
$U$, although the rough estimates of $U_{\textrm{c}}$ 
from Figs.~\ref{fig:poorman}(c) and \ref{fig:poorman}(d) seem consistent 
with those obtained from $m_{s}$.

\begin{figure}
\centering
 \includegraphics[width=0.50\textwidth]{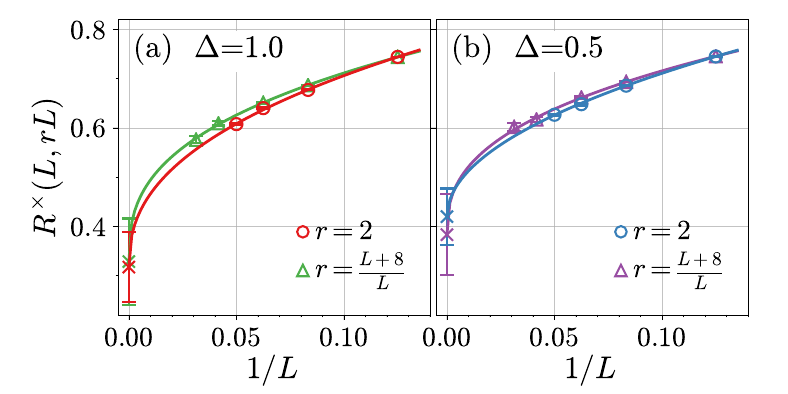}
\caption{\label{fig:Rx}%
$1/L$-extrapolation of crossing points of $R_{m^{2}}(U,L)$
for (a) $\Delta=1.0$ and (b) $\Delta=0.5$.
}
\end{figure}
Based on the previous studies of the honeycomb lattice model~\cite{Meng2010,Sorella2012,Otsuka2013},
it is anticipated that the conventional method to deal directly with the 
order parameters discussed above tends to overestimate the critical points. 
Furthermore, it is 
not trivial to take into account a possible contribution from 
correction terms to the simplest scaling ansatz.
Thus, here we take a more sophisticated approach.
First, we calculate the correlation ratio defined as
$R_{m^{2}}(U,L) = 1 - \frac{S(\bm{K}+\bm{b}/L)}{S(\bm{K})}$,
where $\bm{K}=(\pi,\pi)$ is the AF ordering momentum, 
and $\bm{b}$ is the smallest reciprocal-lattice vector~\cite{Kaul2015}.
Like the Binder ratio,  this quantity has the advantage of being 
size-independent at the critical point when the correction terms are negligible.
Conversely, we can know that the effects of the correction are non-negligible
if we observe that curves of $R_{m^{2}}(U,L)$ as the function of $U$ cross at
different points of $U$ for various $L$. 
Since such a drift is indeed noticed in our results,
we employ the crossing-point analysis;
we determine the crossing points $U^{\times}(L, rL)$ at which the curves of
$L$ and $rL$ cross and extrapolate them to $1/L=0$ assuming the critical behavior of
$U^{\times}(L,rL)=U_{\textrm{c}}+cL^{-(\omega+1/\nu)}$,
where $c$ is a constant, $\omega$ is an effective correction exponent, and
$\nu$ is the correlation-length exponent~\cite{Shao2016}. 
It turns out that this analysis yields reasonable estimates of the critical
points and the exponents to confirm the expected universal nature
of the quantum criticality~\cite{Otsuka2020a}.
Additionally, in Fig.~\ref{fig:Rx}, we show that
$R^{\times}(L,rL)$, the values of $R_{m^{2}}(U,L)$ at the crossing points,
also follow a similar universal behavior of
$R^{\times}(L,rL)=R_{\textrm{c}}^{\times}+dL^{-\omega}$
with  
$R_{\textrm{c}}^{\times}$=0.33(10) [0.40(7)] 
and
$\omega$                 =0.43(13) [0.46(11)] 
for $\Delta$=1.0 (0.5).

\begin{figure}
\centering
 \includegraphics[width=0.24\textwidth]{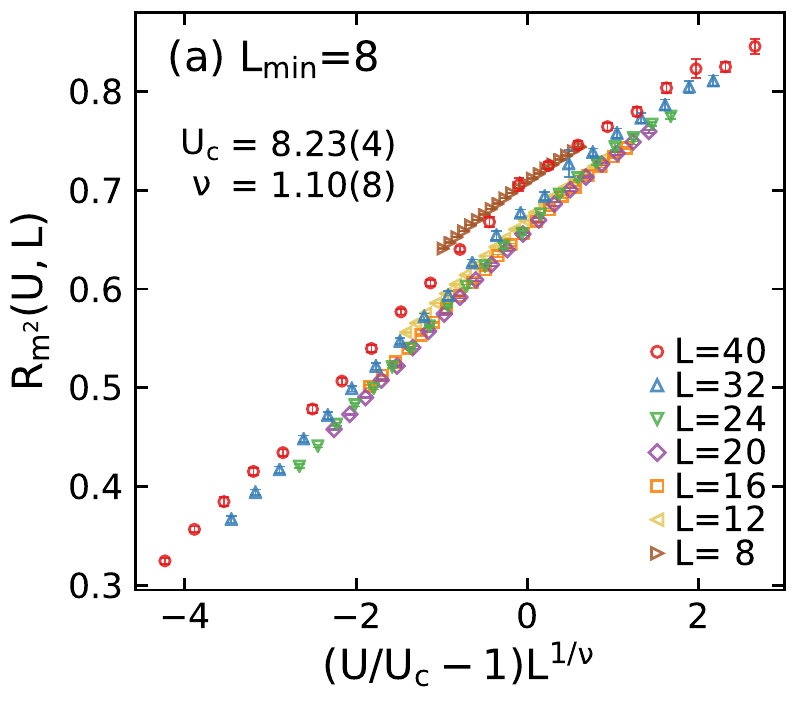}
 \includegraphics[width=0.24\textwidth]{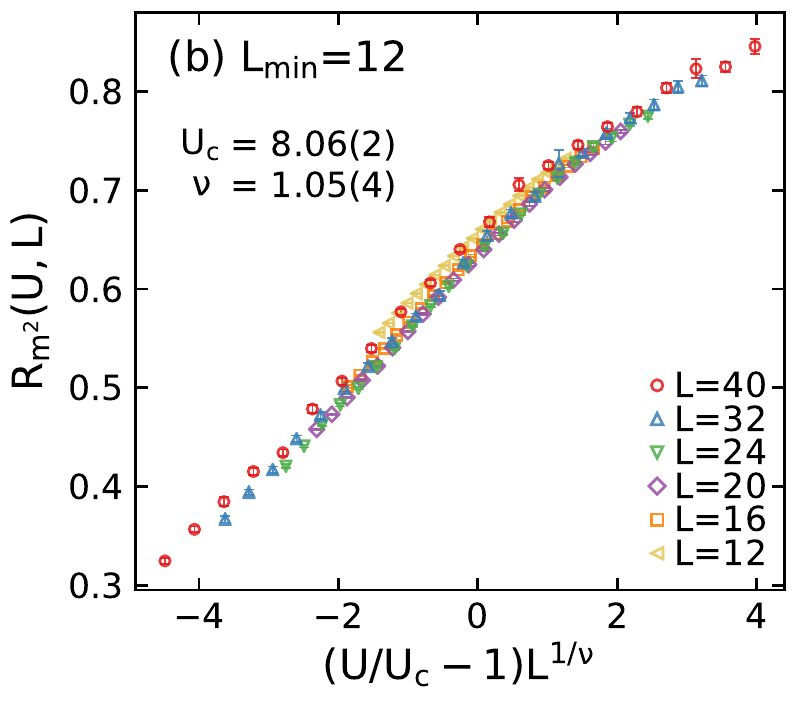}
 \includegraphics[width=0.24\textwidth]{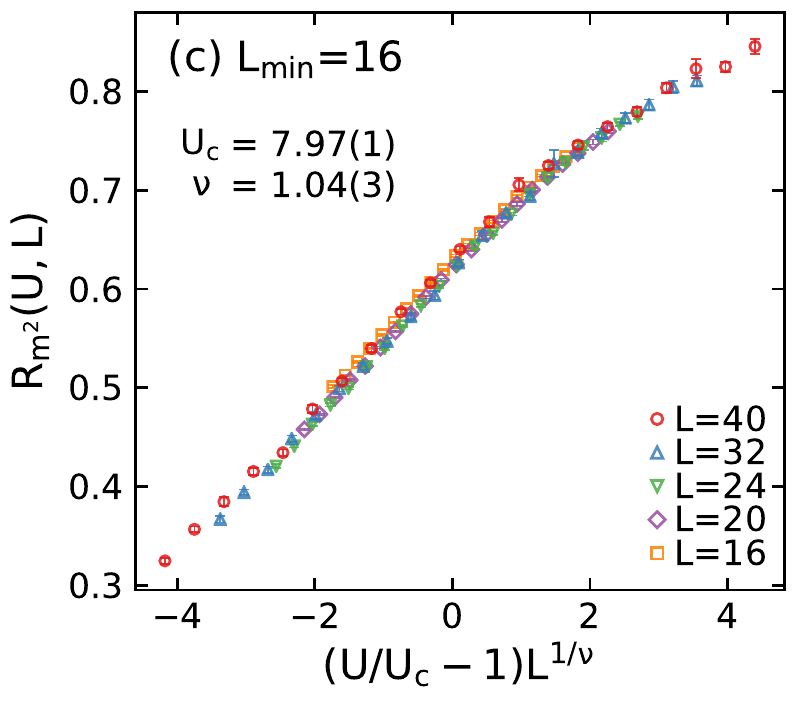}
 \includegraphics[width=0.24\textwidth]{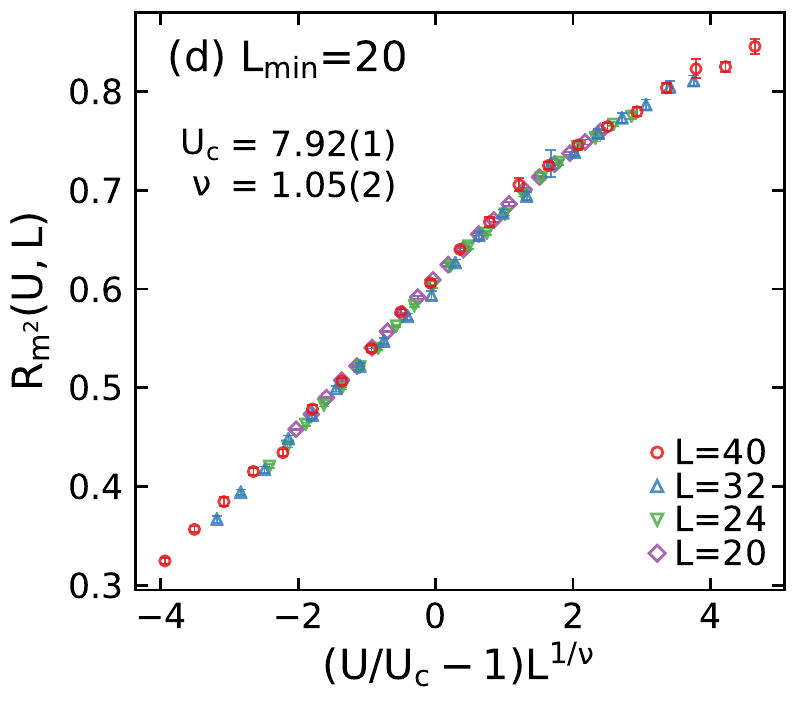}\\
 \includegraphics[width=0.24\textwidth]{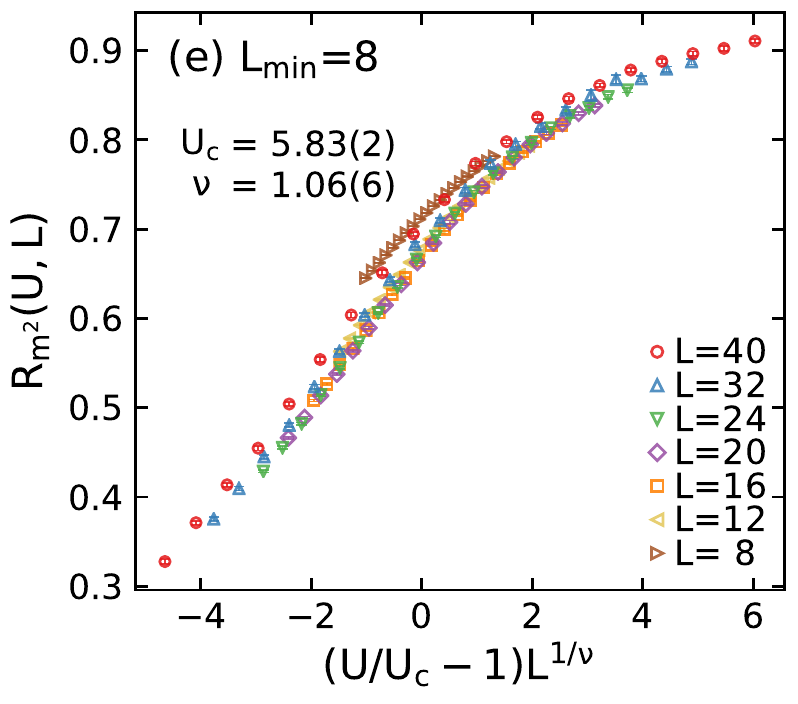}
 \includegraphics[width=0.24\textwidth]{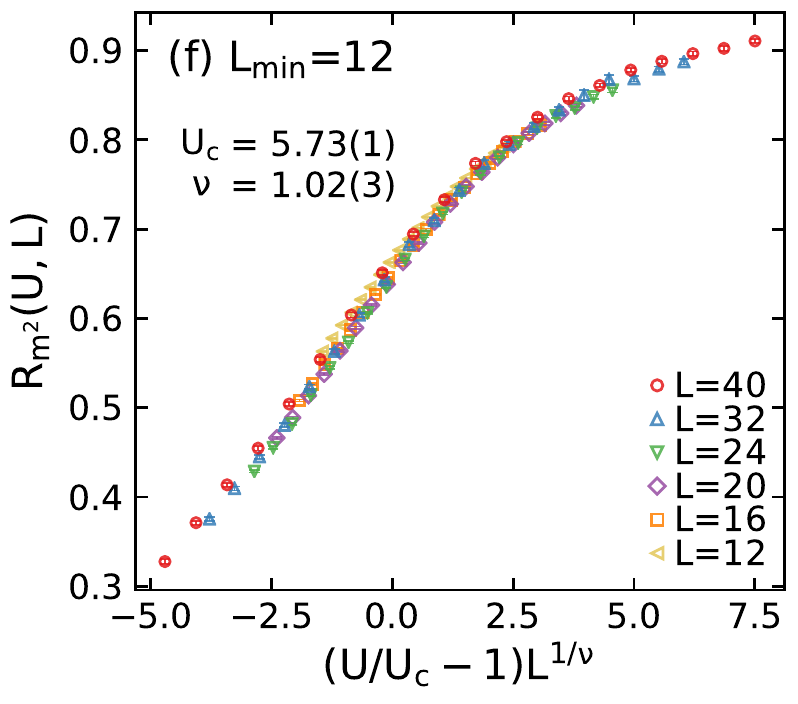}
 \includegraphics[width=0.24\textwidth]{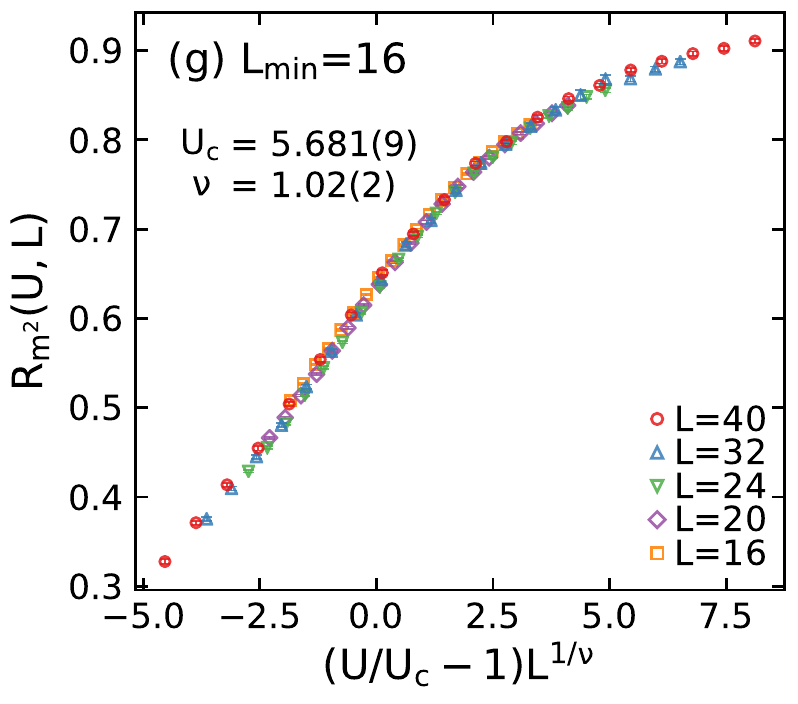}
 \includegraphics[width=0.24\textwidth]{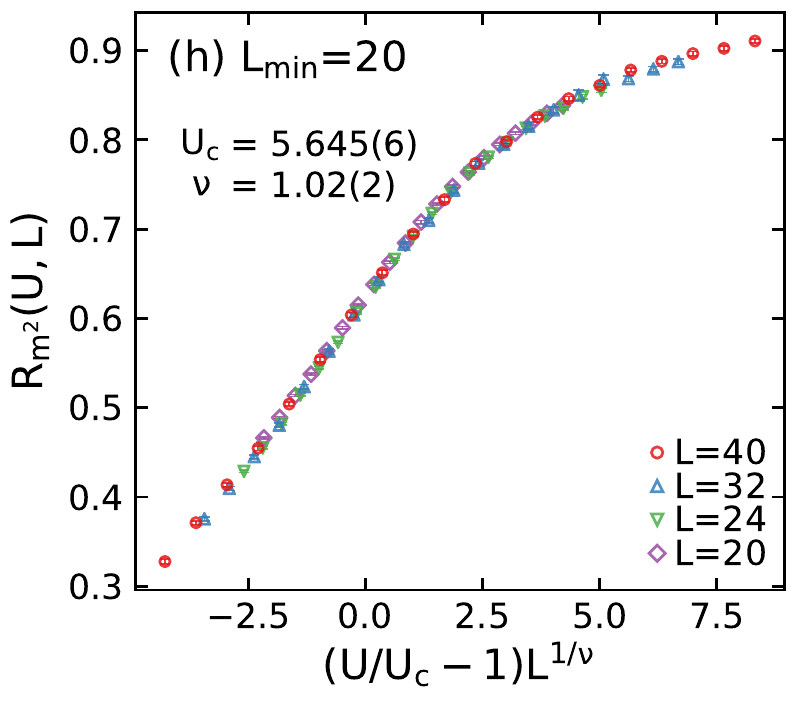}
\caption{\label{fig:collapse_R}%
Data-collapse fits of $R_{m^{2}}(U,L)$ for various $L_{\textrm{min}}$;
(a)-(d) $\Delta=1.0$ and (e)-(f) $\Delta=0.5$.
}
\end{figure}
Finally, we show details of the data-collapse fits of $R_{m^{2}}(U,L)$ 
in Fig.~\ref{fig:collapse_R},
where we collapse the data of $L_{\textrm{min}} \le L \le L_{\textrm{max}}(=40)$
changing $L_{\textrm{min}}$.
Since we utilize the simplest form of the finite-size-scaling ansatz 
without the correction terms, 
$R_{m^{2}}(U,L)=f_{R}(uL^{1/\nu})$, 
where $f_{R}(\cdot)$ denotes the scaling function, 
and $u=(U-U_{\textrm{c}})/U$,
the non-negligible contribution from the correction terms are evident 
for smaller $L_{\textrm{min}}$.
The values of $U_{\textrm{c}}$ and $\nu$ estimated from the collapse fits 
for each $L_{\textrm{min}}$ can be extrapolated to the thermodynamics limit
by plotting the data as a function of $1/L_{\textrm{min}}$, and
the results are consistent with those obtained by the crossing-point analysis
within the error bars.
Readers interested in more details of the critical exponents and comparison 
with analytical results are referred to Ref.~\cite{Otsuka2020a}.

\section{Summary}

We have investigated the square-lattice Hubbard model with a $d$-wave pairing 
field by large-scale auxiliary-field quantum Monte Carlo simulations with the 
aim to revisit the chiral-Heisenberg universality class.
The critical exponents estimated by several ways are overall consistent with 
each other and also with those obtained in the previous studies of the Hubbard
model on the honeycomb lattice and on the square lattice with $\pi$-flux,
suggesting that all these models belong to the same chiral-Heisenberg 
universality class described by the Gross-Neveu model with the same number of 
the fermion components $N=8$.
We also confirm that the anisotropy of the Dirac cones does not affect the
quantum criticality.

\ack
The authors thank
F.~F.~Assaad, T.~Sato, F.~Parisen~Toldin, and Z.~Wang
for valuable comments.
This work has been supported by 
Grant-in-Aid for Scientific Research from MEXT Japan 
(under Grant Nos. JP18K03475, JP18H01183, JP19K23433, JP21H04446, and JP21K03395)
and by PRIN2017 MIUR prot.2017BZPKSZ. 
The numerical simulations have been performed on 
K computer provided by the RIKEN Center for Computational Science (R-CCS) 
through the HPCI System Research project 
(Project IDs: hp170162 and hp170328),
and the HOKUSAI supercomputer at RIKEN (Project IDs: G20006 and Q21525).

\section*{References}
\bibliography{H_BCS-Proc}

\providecommand{\newblock}{}
\begin{thebibliography}{10}
\expandafter\ifx\csname url\endcsname\relax
  \def\url#1{{\tt #1}}\fi
\expandafter\ifx\csname urlprefix\endcsname\relax\def\urlprefix{URL }\fi
\providecommand{\eprint}[2][]{\url{#2}}

\bibitem{Peczak1991}
Peczak P, Ferrenberg A~M and Landau D~P 1991 {\em Phys. Rev. B\/} {\bf 43} 6087

\bibitem{Holm1993}
Holm C and Janke W 1993 {\em Phys. Rev. B\/} {\bf 48} 936

\bibitem{Campostrini2002}
Campostrini M, Hasenbusch M, Pelissetto A, Rossi P and Vicari E 2002 {\em Phys.
  Rev. B\/} {\bf 65} 144520

\bibitem{Wang2006}
Wang L, Beach K~S~D and Sandvik A~W 2005 {\em Phys. Rev. B\/} {\bf 73} 014431

\bibitem{Wenzel2009}
Wenzel S and Janke W 2009 {\em Phys. Rev. B\/} {\bf 79} 014410

\bibitem{Shao2016}
Shao H, Guo W and Sandvik A~W 2016 {\em Science\/} {\bf 352} 213

\bibitem{Ma2018}
Ma N, Weinberg P, Shao H, Guo W, Yao D~X and Sandvik A~W 2018 {\em Phys. Rev.
  Lett.\/} {\bf 121} 117202

\bibitem{Sorella1992}
Sorella S and Tosatti E 1992 {\em Europhys. Lett.\/} {\bf 19} 699

\bibitem{Paiva2005}
Paiva T, Scalettar R~T, Zheng W, Singh R~R~P and Oitmaa J 2005 {\em Phys. Rev.
  B\/} {\bf 72} 085123

\bibitem{Meng2010}
Meng Z~Y, Lang T~C, Wessel S, Assaad F~F and Muramatsu A 2010 {\em Nature\/}
  {\bf 464} 847

\bibitem{Sorella2012}
Sorella S, Otsuka Y and Yunoki S 2012 {\em Sci. Rep.\/} {\bf 2} 992

\bibitem{Otsuka2013}
Otsuka Y, Yunoki S and Sorella S 2013 {\em J. Phys. Conf. Ser.\/} {\bf 454}
  012045

\bibitem{Herbut2006}
Herbut I~F 2006 {\em Phys. Rev. Lett.\/} {\bf 97} 146401

\bibitem{Assaad2013}
Assaad F~F and Herbut I~F 2013 {\em Phys. Rev. X\/} {\bf 3} 031010

\bibitem{Rosenstein1993}
Rosenstein B, Kovner A, {Hoi-Lai Yu} and Kovner A 1993 {\em Phys. Lett. B\/}
  {\bf 314} 381

\bibitem{ParisenToldin2015}
{Parisen Toldin} F, Hohenadler M, Assaad F~F and Herbut I~F 2015 {\em Phys.
  Rev. B\/} {\bf 91} 165108

\bibitem{Otsuka2016}
Otsuka Y, Yunoki S and Sorella S 2016 {\em Phys. Rev. X\/} {\bf 6} 011029

\bibitem{Otsuka2002}
Otsuka Y and Hatsugai Y 2002 {\em Phys. Rev. B\/} {\bf 65} 073101

\bibitem{Otsuka2014}
Otsuka Y, Yunoki S and Sorella S 2014 {\em JPS Conf. Proc.\/} {\bf 3} 013021

\bibitem{Otsuka2020a}
Otsuka Y, Seki K, Sorella S and Yunoki S 2020 {\em Phys. Rev. B\/} {\bf 102}
  235105

\bibitem{Ixert2014}
Ixert D, Assaad F~F and Schmidt K~P 2014 {\em Phys. Rev. B\/} {\bf 90} 195133

\bibitem{Hatsugai2006}
Hatsugai Y, Fukui T and Aoki H 2006 {\em Phys. Rev. B\/} {\bf 74} 205414

\bibitem{Blankenbecler_PRD1981}
Blankenbecler R, Scalapino D~J and Sugar R~L 1981 {\em Phys. Rev. D\/} {\bf 24}
  2278

\bibitem{Hirsch_PRB1985}
Hirsch J~E 1985 {\em Phys. Rev. B\/} {\bf 31} 4403

\bibitem{White_PRB1989}
White S~R, Scalapino D~J, Sugar R~L, Loh E~Y, Gubernatis J~E and Scalettar R~T
  1989 {\em Phys. Rev. B\/} {\bf 40} 506

\bibitem{Seki2019}
Seki K, Otsuka Y, Yunoki S and Sorella S 2019 {\em Phys. Rev. B\/} {\bf 99}
  125145

\bibitem{Kaul2015}
Kaul R~K 2015 {\em Phys. Rev. Lett.\/} {\bf 115} 157202

\end{thebibliography}


\end{document}